\documentclass[journal=jacsat,manuscript=article]{achemso}

\usepackage[version=3]{mhchem} 
\usepackage{hyperref}

\newcommand{\form}{$\Delta E_{\mathrm{vac}}^f$}
\newcommand{\ads}{$\Delta E_{\mathrm{ads}}^f$}
\newcommand{\bind}{$\Delta E_{\mathrm{b}}$}
\newcommand{\etdsub}{$E_{\mathrm{2D+S}}$}
\newcommand{\etdtd}{$E_{\mathrm{3D}}$}
\newcommand{\ntdtd}{$N_{\mathrm{3D}}$}
\newcommand{\etd}{$E_{\mathrm{2D}}$}
\newcommand{\ntd}{$N_{\mathrm{2D}}$}
\newcommand{\esub}{$E_{\mathrm{S}}$}
\newcommand{\ehull}{$E_{\mathrm{above\ hull}}$}
\newcommand{\sample}{SbTeI}

\author{Tara M. Boland}
\affiliation[Denmark] {Computational Atomic-Scale Materials Design (CAMD), Technical University of Denmark, DK-2800 Kgs. Lyngby, Denmark}
\altaffiliation{Authors contributed equally}
\author{Rachel Gorelik}
\affiliation[SEMTE] {School for Engineering of Matter, Transport and Energy, Arizona State University, Tempe, AZ, USA}
\altaffiliation{Authors contributed equally}
\author{Arunima K. Singh}
\affiliation[Physics] {Department of Physics, Arizona State University, Tempe, AZ, USA}
\email{arunimasingh@asu.edu}

\title[Heterostructures]
  {Fundamental Factors Governing  Stabilization of Janus 2D-Bulk Heterostructures with Machine Learning}

\abbreviations{}
\keywords{American Chemical Society, \LaTeX}

\begin{document}


\begin{abstract}
The more-than-6000 2D materials predicted thus far provide a huge combinatorial space for forming functional heterostructures with bulk materials, with potential applications in nanoelectronics, sensing, and energy conversion. In this work, we investigate nearly 1000 heterostructures, the largest number of heterostructures thus far, of 2D Janus and bulk materials' surfaces using \emph{ab initio} simulations and machine learning (ML) to deduce the structure-property relationships of the complex interfaces in such heterostructures. We first perform van der Waals-corrected density functional theory simulations using a high-throughput computational framework on 51 Janus 2D materials and 19 metallic, cubic phase, elemental bulk materials that exhibit low lattice mismatches and low coincident site lattices. The formation energy of the resultant 1147 Janus 2D-bulk heterostructures were analyzed and 828 were found to be thermodynamically stable. ML models were trained on the computed data, and we found that they could predict the binding energy  and $z$-separation of 2D-bulk heterostructures with root mean squared errors (RMSE) of 0.05 eV/atom and 0.14 \AA, respectively. The feature importance of the models reveals that the properties of the bulk materials dominate the heterostructures' energies and interfacial structures heavily. These findings are in-line with experimentally observed behavior of several well-known 2D materials-bulk systems. The data used within this paper is freely available in the \href{http://hydrogen.cmd.lab.asu.edu/data}{\emph{Ab Initio} 2D-Bulk Heterostructure Database} (aiHD). The fundamental insights on 2D-bulk heterostructures and the predictive ML models developed in this work could accelerate the application of thousands of 2D-bulk heterostructures, thus stimulating research within a wide range of electronic, quantum computing, sensing, and energy applications.
\end{abstract}

\section{Introduction}

Two-dimensional (2D) materials are a class of atomically thin materials, ranging from single-to-few atomic layers in thickness, with a wide variety of material properties resulting from quantum confinement effects.

Many 2D materials have been experimentally synthesized since the discovery of graphene in 2004~\cite{Novoselov2004}, with various methods subsequently explored for tuning their properties, such as strain engineering, doping, and forming heterostructures (either with other 2D materials or with bulk materials)~\cite{singh2015al2o3, Singh2014a, houlong2017doping, smirman2017, Paul2017, Singh2014, Singh2015, Singh2015a}. Prior works have studied 2D-2D heterostructures but there has been little exploration of the potentially differing phenomena of 2D-bulk interfaces~\cite{pakdel2024high, Zhang2023, Rhodes2019, Akiyama2021, Ozcelik2016, Geim2013, Willhelm2022, zeng2018novel}. Heterostructures of 2D and bulk materials could provide a huge combinatorial space for the formation of functional heterostructures that will facilitate the low-barrier integration of novel 2D materials with existing bulk technologies such as nanoelectronics, quantum computing, chemical sensing, and energy conversion.

More recently, the high-throughput \emph{ab initio}/theory-based approach to materials discovery has led to significant advancements in 2D materials--i.e. the prediction of over 7000 2D materials and the experimental realization of over 5 \% of these materials.~\cite{c2db, Ashton2017, Ozcelik2016, Choudhary2017, Zhang2019, Campi2022} Crystal structure and properties of 2D materials have been systematically curated in more than five open-source databases~\cite{c2db, Ashton2017, Ozcelik2016, Choudhary2017, Zhang2019, Campi2022}, guiding future studies towards further experimental realization and application of 2D materials. Similarly, there are currently more than 10 databases that curate the structural, electronic, mechanical, and optical properties of bulk materials obtained from \emph{ab initio} methods such as density functional theory (DFT) that also facilitate machine learning (ML) and data-driven sciences.~\cite{aflow, Jain2013, oqmd, ICSD} However, there are no systematic high-throughput simulations or open-source databases of 2D-bulk heterostructures. 

 \begin{figure}[h!]
        \centering
        \includegraphics[width=3.33in]{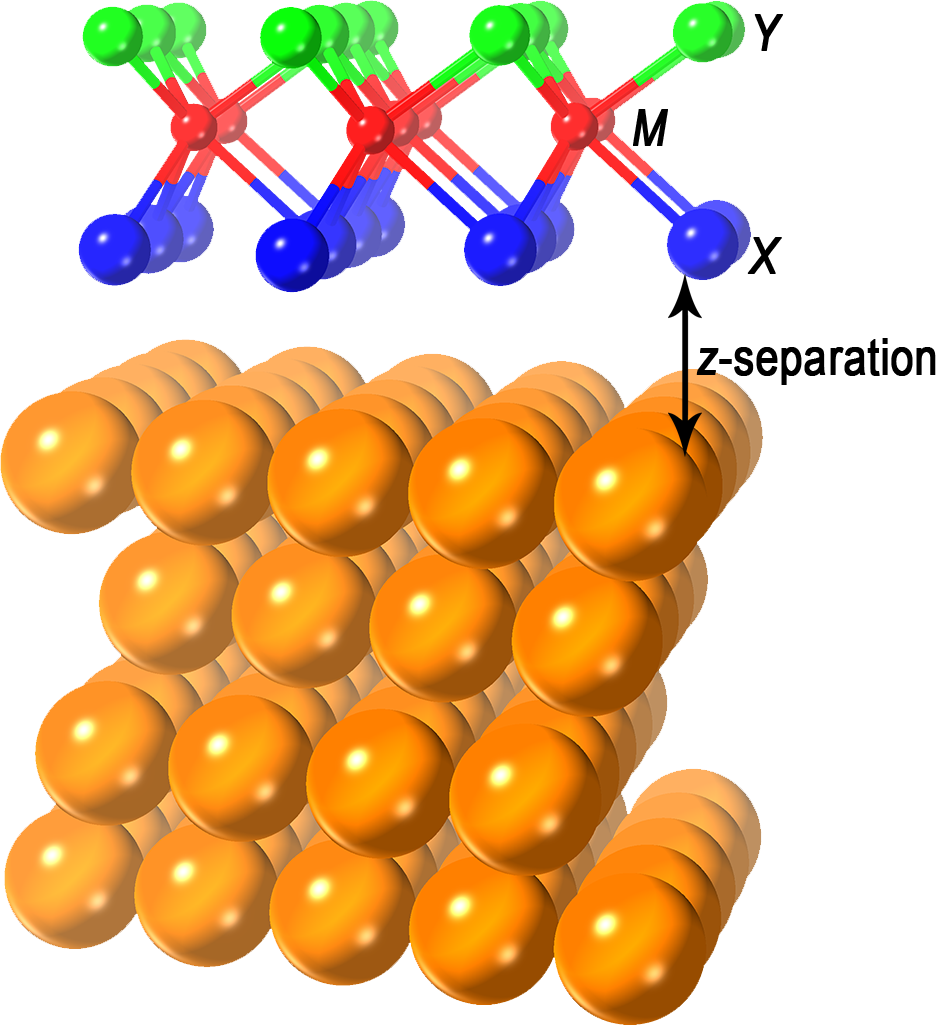}
        \caption[Structure Model of Y-M-X Janus 2D Material with Substrate]{$1H$-Janus 2D-bulk heterostructure model showing the orientation of the 2D material on top of the bulk surface. The Y-M-X represents the atomic layers of the Janus 2D material.}
        \label{fig:janus_model}
    \end{figure}

In this report, we introduce the \href{http://hydrogen.cmd.lab.asu.edu/data}{\emph{Ab Initio} 2D-Bulk Heterostructure Database} (aiHD), the first freely available database of \emph{ab initio}-calculated data on 2D-bulk heterostructures and their post-adsorbed properties. To construct, optimize, and analyze the 2D-bulk heterostructures, we use \href{https://github.com/cmdlab/Hetero2d}{Hetero2d}~\cite{hetero2d}, a high-throughput computational synthesis package designed for computing heterostructure properties within the DFT framework. We explore the interaction of metastable 2D materials with various bulk materials' surfaces through DFT calculations for a set of 51 Janus 2D materials and 19 elemental bulk substrates in lattice- and symmetry-matched 2D-bulk heterostructures. Janus 2D materials with formula Y-M-X -- which are 3-atomic-layers-thick, with each atomic layer consisting of a distinct element -- exhibit unique properties, including finite out-of-plane dipole moments, in-plane piezoelectricity~\cite{Riis2019, Zhang2020}, and magnetic skyrmions ~\cite{Hou2022}. From our computed dataset of 1147 Janus 2D-bulk pairs, we identify 828 2D-bulk configurations that stabilize the metastable Janus 2D materials. ML random forest regression models developed with this \textit{ab-initio}-computed data predict the binding energy and $z$-separation distance for 2D-bulk heterostructures with low root mean squared errors (RMSE) of 0.047 eV/atom and 0.139 \AA\ , respectively. Through these models, the bulk material's electronegativity and surface energy are each found to be the leading fundamental factor that provides bulk-induced stabilization of a 2D-bulk pair's binding energy and $z$-separation distance, respectively. 

This work makes a major advancement in the understanding of heterostructures of 2D materials and bulk materials. It is poised to strongly guide the experimental realization and utilization of the thousands of previously unsynthesized 2D materials for a wide variety of applications such as photocatalysis, electronics, and batteries. The ML models developed in this work will significantly reduce the computational costs of extending such studies to other 2D-bulk heterostructures. Furthermore, the aiHD and the ML models can facilitate targeted searches of large combinatorial spaces of heterostructures for the exploration of stable 2D-bulk pairs and their corresponding properties.

\section{Results and Discussion}

\subsection{Screening Parameters and Materials Selection}
    \subsubsection{Materials Selection}

    The 51 Janus 2D materials, Y-M-X, (Y=[S, Se, Te], M=[As, Bi, Cr, Hf, Mo, Nb, Sb], and X=[Br, Cl, I, Se, Te]) were obtained from the Computational 2D Materials Database (C2DB)~\cite{c2db}, which contains the DFT-computed structures of these Janus 2D materials which exhibit a diverse range of predicted properties. Two different Janus 2D polytypes, 1$H$ and 1$T$, were considered for each Janus 2D material; an example of the 1$H$ polytype structure is shown in Figure \ref{fig:janus_model}, while a 1$T$ polytype shifts the $x$ and $y$ coordinates of the  X and Y atoms with respect to each other\cite{Paul2017}. The polytype, formation energy, C2DB ID, and spacegroup for each Janus 2D material are listed in Supplementary Information (SI) Table 1.
    
    The 3D phase of each 2D material was obtained from the Materials Project (MP)~\cite{Ong2013} database using the (1) lowest \ehull\ and (2) same elemental composition as the 2D material. \ehull\ is the energy of decomposition of a material into the set of most stable materials at a given chemical composition. SI Table 2 lists the Materials Project ID and spacegroup for each 3D bulk phase(s) and discussion for each composition-matched 3D bulk phase(s).
    
    For the bulk materials that act as substrates for the 2D Janus materials, this work selected metallic, cubic-phase single-elements from the MP database with valid ICSD IDs~\cite{ICSD} (i.e., previously experimentally synthesized), and possessing the lowest \ehull\ . To reduce computational costs, only the 19 bulk materials optimized in our previous study~\cite{hetero2d} were considered, terminated on the (111) surface: Al, Au, Ag, Cu, Hf, Ir, Mn, Nd, Ni, Pd, Re, Rh, Sc, Sr, Te, Ti, V, Y, and Zr. No other low-index Miller planes were found during the lattice-matching step to the aforementioned 2D materials. SI Table 3 lists the Materials Project ID, surface energy,\cite{hetero2d} and spacegroup for each bulk material.
    
    \subsubsection{Heterostructure Screening Criteria}

\begin{figure}[h]
        \centering
        \includegraphics[width=3.33 in]{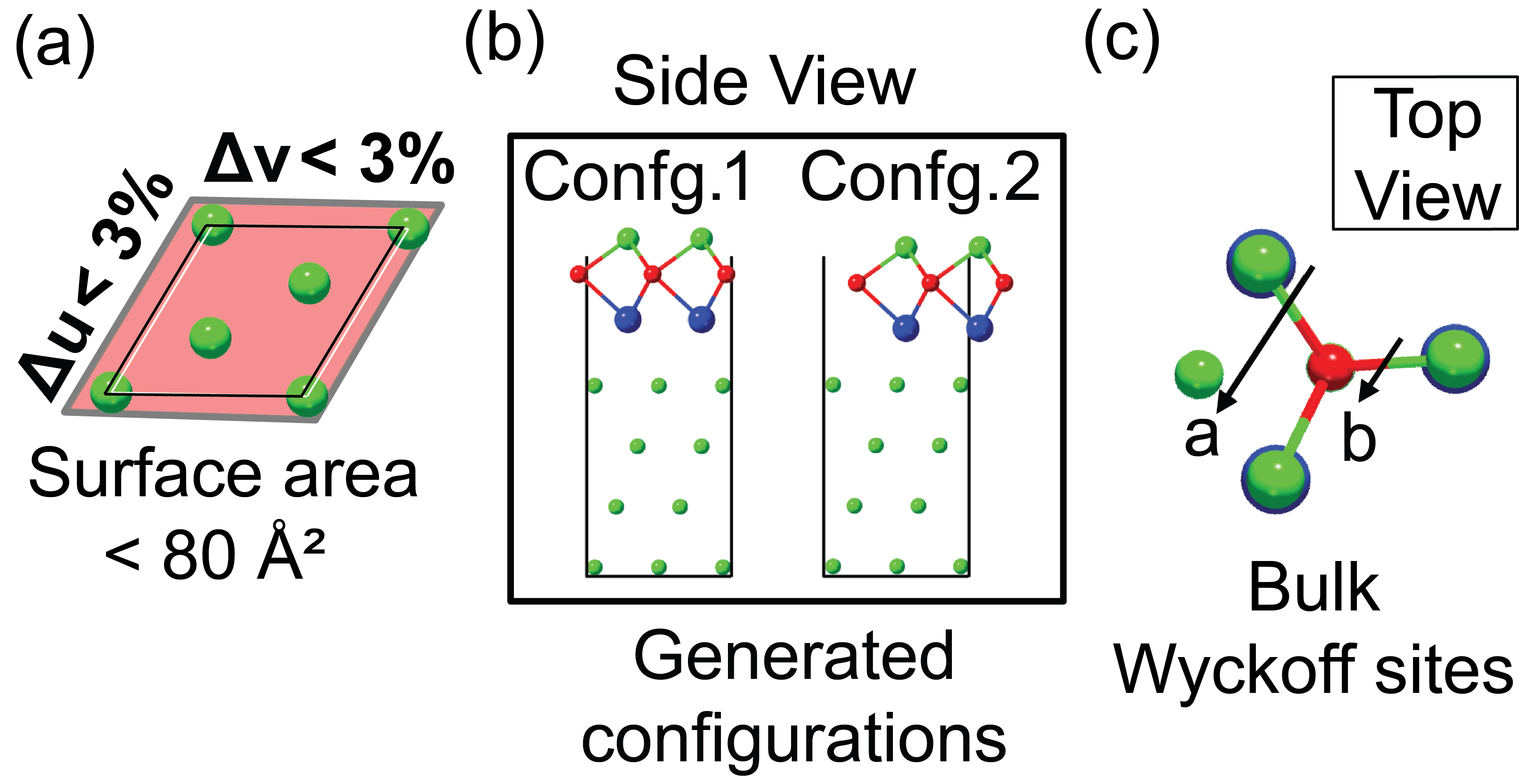}
        \caption{(a) Schematic illustrating the criteria for creating heterostructures, (b) the 2D-bulk heterostructure configurations from the top and side, where (c) the top view denotes how the 2D material was aligned on top of the bulk surface using the Wyckoff sites.}
        \label{fig:hiface_criteria}
    \end{figure}

    A maximum heterostructure surface area of $\gamma_{SA} < 80$ \AA$^2$ and maximum applied strain on the 2D material of $\sigma_{2D} < 3$~\% were applied as constraints for the lattice-matching algorithm~\cite{Zur1984, Mathew2016} (see Figure \ref{fig:hiface_criteria}). The algorithm identified a total of 438 Janus 2D-bulk material pairs, all with the (111) bulk surface termination. A total of 1147 heterostructures were obtained in all when considering the two to three heterostructure configurations generated for each 2D-bulk pair. The \textit{Hetero2d} packages finds all available configurations for a given pair by enumerating the high-symmetry points of the 2D and bulk materials and aligning the 2D material and bulk high-symmetry points less than 4 \AA\ apart, resulting in unique configurations. SI Table 4 lists all the pairs between bulk single-element materials and Janus 2D materials that produce heterostructures with the above constraints.  

\subsection{DFT Methods}
All DFT calculations were performed using methods and parameters as described in detail in Ref. \citep{hetero2d}. As previously demonstrated there, these parameters and methods were optimized for accurate descriptions of the 2D-bulk heterostructures. The DFT relaxation calculations were performed using the projector-augmented wave method as implemented in the plane-wave code VASP~\cite{Kresse1, Kresse2, Kresse3, Kresse4, Kresse5}, and the vdW interactions between the 2D materials and bulk materials were modeled using the vdW–DF~\cite{Rydber2003} functional with the optB88~\cite{Klimes2011} exchange functional. A cutoff energy of 520 eV was used for all calculations. The default $k$-point grid density was automated using pymatgen~\cite{Ong2013} routines to 20 $k$-points/unit length by taking the nearest integer value after multiplying $\frac{1}{a}$ and $\frac{1}{b}$ by 20. These settings were sufficient to converge all calculations to a total force per atom of less than 0.02 eV/\AA. 

The density of states (DOS) and charge density difference~\cite{Tang2009, Henkelman2006} calculations were performed in this work for select heterostructures and were automated using the \textit{CMDLElectronicSet} available in the \textit{Hetero2d} package. The $k$-point mesh grid used to calculate the electronic properties was automated by setting the reciprocal density to 200 with \textit{pymatgen}'s automatic density by volume. The DOS grid was set such that the number of sampling points had an energy spacing of 0.05 eV between each point. The benchmark case, taken with energy spacing of 0.1 eV, 0.05 eV, and 0.01 eV, showed the band gap was converged to within 0.019 eV. The fine grid for the charge density was set to 0.03 \AA\, as this value provided a reasonable compromise between accuracy and computational cost, $\sim$2.3x the default grid spacing. The benchmarking done with grid spacing of 0.1 eV, 0.05 eV, 0.03, and 0.01 eV, showed the Bader charges were converged to within $\sim$0.004 electrons per atom. The charge density difference plots demonstrated little change with increased grid spacing.

\section{Machine Learning Methodology} 

    Random forest regression (RFR) models were developed in this work to predict the binding energy, \bind, and the $z$-separation distance at the interface of the heterostructures using the \textit{scikit-learn} code~\cite{Pedregosa2011, sklearn_api}. RFR models have been successfully applied to smaller datasets with less than 1000 data points~\cite{jain2024machine, pilania2021machine, biswas2024incorporating}. The hyperparameters were tuned for each model using \textit{GridSearchCV}, with optimized hyperparameters listed in SI Section Machine Learning Convergence Testing. The \bind\ and $z$-separation RFR models were trained with a $K$-fold cross-validation value of $K$=8 and $K$=5, respectively, and the optimized models for \bind\ and $z$-separation employed 150 and 500 decision trees. These parameters were found to provide reasonable accuracy and computational speed for training and testing.

   \begin{figure}[!h]
        \centering
        \includegraphics[width=3.33in]{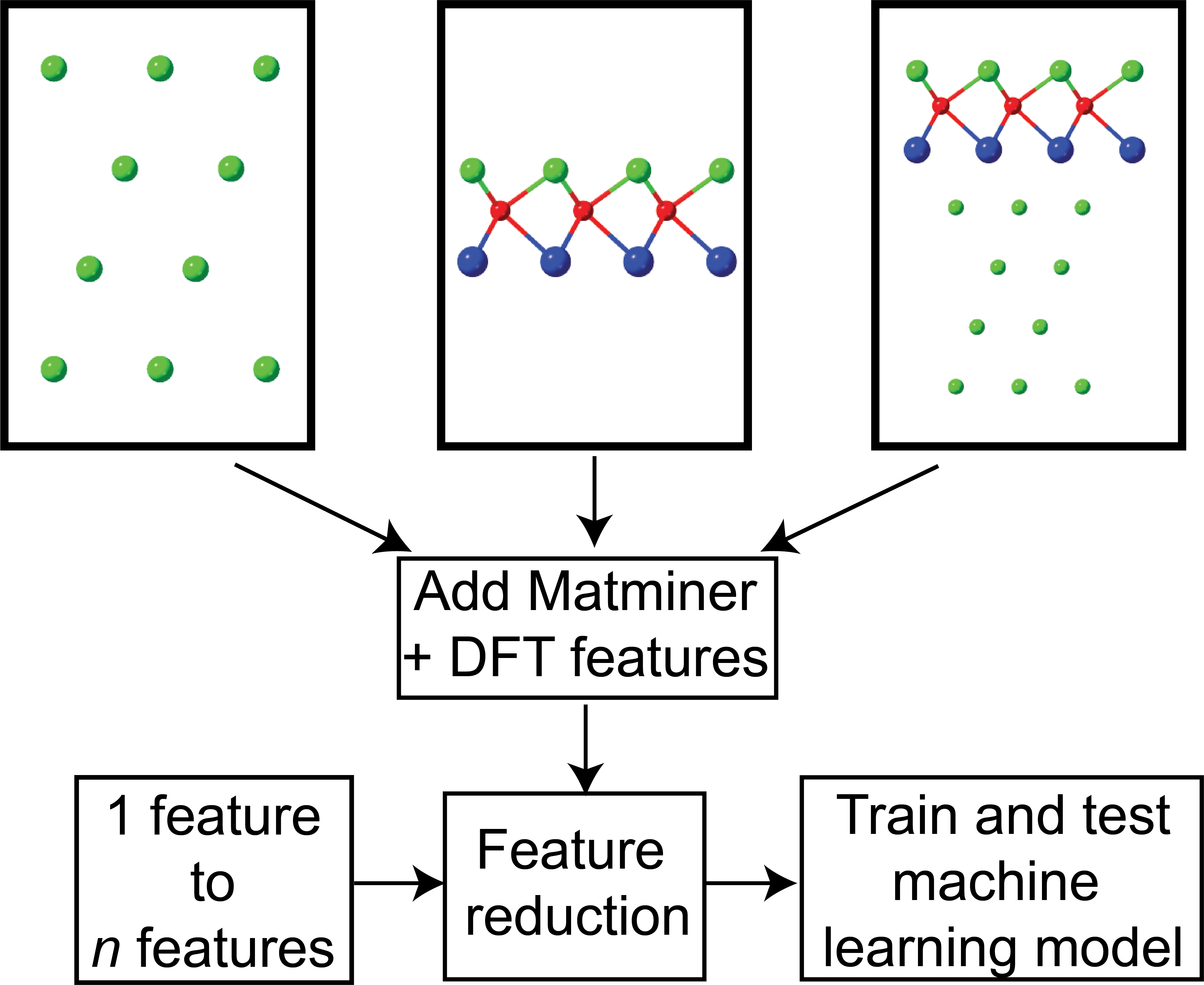}
        \caption{Schematic diagram of the steps taken to prepare the data set and features to perform machine learning.}
        \label{fig:ML_workflow}
    \end{figure}

     Figure~\ref{fig:ML_workflow} shows a schematic of the steps taken to prepare the data set and the features to perform ML. To generate feature descriptors for the heterostructure and its constituent materials, three resources were used: (a) material properties queried from C2DB (b) calculation of \textit{ab-initio} information using DFT, and (c) the Matminer package~\cite{Ward2018}. The Matminer provided both compositional-based features from the MP database as well as structural-based features from this study's DFT-relaxed 2D-bulk heterostructures. C2DB was used to obtain 2D materials-related features. A more detailed description of features can be found in SI. The data set consisted of the lowest-energy configuration for each of the 438 Janus 2D-bulk heterostructure pairs optimized using vdW-corrected DFT.

   We obtained a total of 155 features for each 2D-bulk heterostructure. A systematic and rational feature reduction was then performed to obtain optimal features for the RFR models. The features removed were either low-variance or highly correlated to other features. Through test models, low-variance features were defined as those features with numeric values of a percent occurrence greater than 80 \%. The highly-correlated features were determined using \textit{pandas} correlation matrix function \textit{corr}. Threshold values from 10 \%-95 \% were searched over successively to find the optimal value which improves or maintains the test model's performance. A threshold value of 65 \% was found to improve the model performance, which was measured using the R$^2$, RMSE, and MAE for the range of threshold values. Finally, additional unnecessary features were removed using the recursive feature elimination with cross-validation (RFECV) method implemented in \textit{scikit-learn}.
   
   The feature reduction resulted in a total of 15 remaining features for the \bind\ RFR model from the 155 starting feature set. Similarly, the $z$-separation model was reduced to a total of 10 features.  A more detailed description of features, feature reduction, performance metrics for each model, and the hyperparameter tuning strategy can be found in SI Section Machine Learning Convergence Testing.

\subsection{Visualization of the Data Set and Trends}

    \subsubsection{Energetic Stability of Janus 2D Materials}

    In order to assess the thermodynamic stability of the 2D-bulk heterostructures, we define an adsorption energy, \ads = \form\ - \bind~\cite{Singh2015}, consisting of the 2D formation energy and 2D-bulk binding energy. The formation energy of the 2D material \form\ is given by $\Delta E_{\mathrm{vac}}^f = E_{\mathrm{2D}}/N_{\mathrm{2D}} - E_{\mathrm{3D}}/N_{\mathrm{3D}}$, where \etd\ is the energy of a 2D material in vacuum, \etdtd\ is the energy of the 3D counterpart of the 2D material, and \ntd\ and \ntdtd\ are the number of atoms in the unit cell of their 2D and 3D counterparts, respectively. The binding energy \bind\ of the 2D-bulk heterostructure pair is given by $\Delta E_{\mathrm{b}} = (E_{\mathrm{2D}} + E_{\mathrm{S}} - E_{\mathrm{2D+S}} )/N_{\mathrm{2D}}$, where \etdsub\ is the energy of the 2D material adsorbed on the surface of a bulk substrate, \esub\ is the energy of the bulk slab, \etd\ is the energy of the freestanding 2D material, and \ntd\ is the number of atoms in the unit cell of the 2D material.

    When \ads\ (i.e. the difference between \form\ and \bind\ ) is less than zero, the bulk material stabilizes the 2D material. The lower \ads\ is, the higher the heterostructure's stability (while, conversely, the lower the \bind\, the lower the stability). Of the 438 2D-bulk pairs, 318 pairs (a total of 828 heterostructure configurations) have negative adsorption energies, indicating stabilization of the 2D material by the bulk slab. 
    
\begin{figure}[!h]
        \centering
        \includegraphics[width=3.33in]{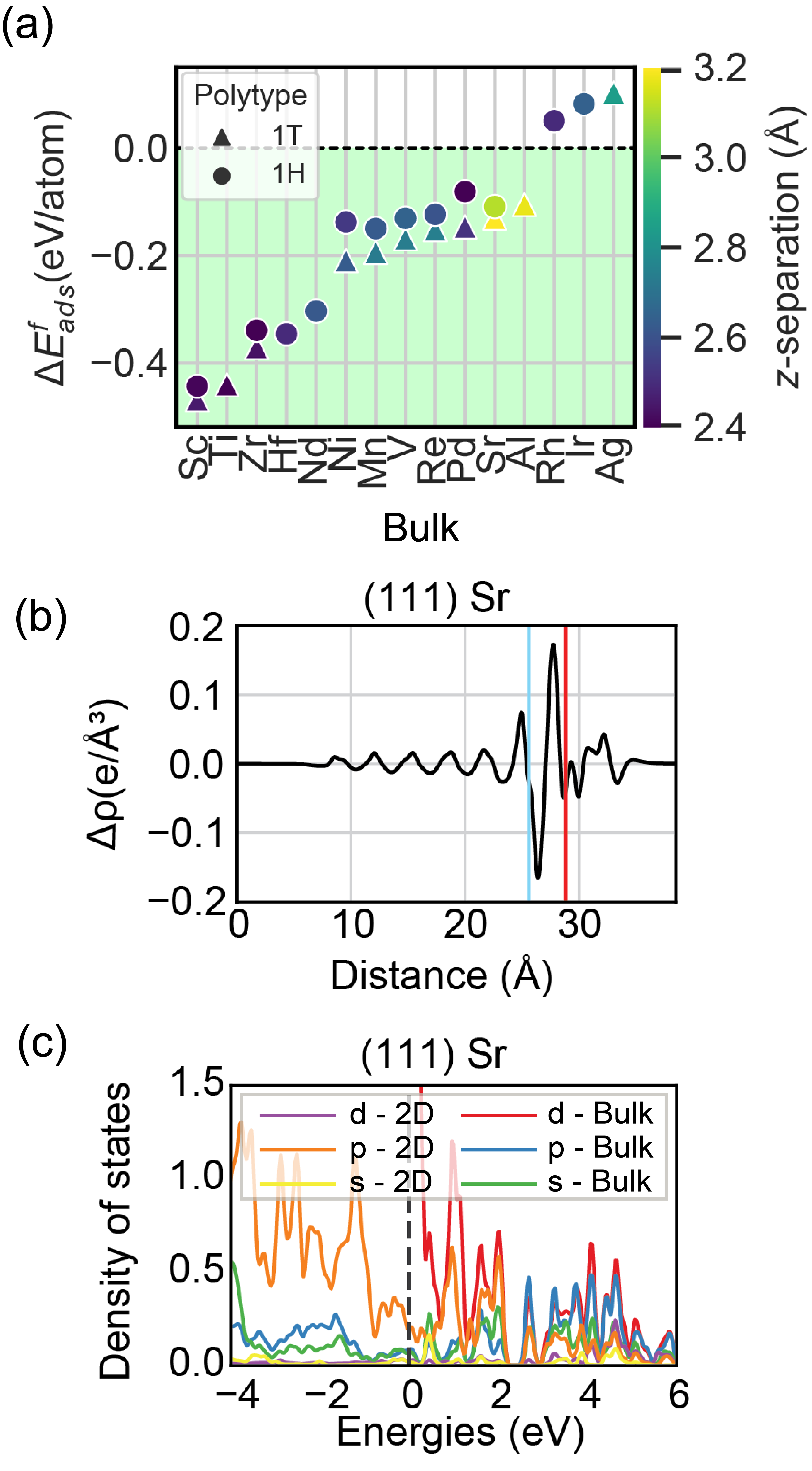}
        \caption{(a) Adsorption formation energy, \ads, of $1T$- (triangle markers) and $1H$-\sample\ (circle markers) on the lattice- and symmetry-matched bulk surfaces. The bulk materials are noted on the $x$-axis, and the \ads\ for the lowest energy heterostructure configuration is shown on the $y$-axis. The green shaded region indicates when the bulk material stabilizes the 2D material. The color map corresponds to the $z$-separation distance between the 2D materials and the bulk surface. For the Sr(111)-$1H$-\sample\, (b) shows the $z$-projected charge density difference, where the vertical blue and red lines are average $z$-positions of the top-most atomic layer of the bulk surface and the bottom-most atomic layer of the 2D material, respectively. The positive values represent electron accumulation and negative values represent electron density depletion with respect to the isolated 2D material and  bulk slab (c) shows the site and orbital projected DOS for the atoms at the interface.}
        \label{fig:eads_janus}
    \end{figure}

    Figure \ref{fig:eads_janus}(a) shows the \ads\ for the $1T$ and $1H$ phases of \sample\ on 15 bulk substrates. The degree to which the 2D material is stabilized varies depending on the bulk material type. Previous reports~\cite{Singh2014b, singh2015al2o3} have indicated that larger \ads\ are more consistent with ionic/covalent interactions, while values closer to zero are more consistent with vdW type interactions~\cite{ singh2015al2o3,Singh2014, Singh2015a, hetero2d}. vdW interactions ($\sim$0.03 eV/atom or less) are typically an order of magnitude smaller than ionic/covalent bonds ($\sim$0.2 eV/atom)~\cite{Deng2017}. Excluding Rh, Ir, and Ag, the (111) surface of all the bulk materials stabilize the \sample\ 2D material. The strength of their interaction varies from relatively weak (near zero) to large values of $\sim$-0.4 eV/atom. 

    The $z$-separation of a given heterostructure is defined in this work as $z_{avg}^{2D} - z_{avg}^{bulk}$, where $z_{avg}^{2D}$ is the average $z$-coordinate of the bottom-most 2D layer, and $z_{avg}^{bulk}$ is the average $z$-coordinate of the topmost layer of atoms in the bulk material. The heterostructure pairs with large negative adsorption energies are expected to exhibit smaller $z$-separation distances at the 2D-bulk interface due to the stronger binding between the two surfaces. Indeed, this trend is observed for the systems studied in the work, with the color map in Figure \ref{fig:eads_janus}(a) displaying the correlation between decreasing \ads\ and $z$-separation distance for \sample\ . 

    To further understand the effect of the bulk material on the 2D material and to provide insight into the nature of the bonding,  DOS calculations and Bader charge\cite{Henkelman2006} analysis were performed on a case study of the $1H$-\sample\ -Sr(111) heterostructure. Figure \ref{fig:eads_janus}(b) shows the $z$-projected charge density difference for the system, as a graphical means of determining the charge distribution from the interactions between the 2D material and the bulk slab. The charge density difference requires three calculations of charge density: (1) the combined system, (2) the isolated 2D material, and (3) the isolated substrate slab in which the atomic position of each atom are frozen in each case. The vertical blue line represents the average $z$-position of the top-most atomic layer of the substrate surface, and the red line is the average $z$-position of the bottom-most atomic layer of the 2D material. The charge density difference in Figure \ref{fig:eads_janus}(b) shows significant redistribution of the charge density at the interface due to the interactions between the 2D material and bulk surface. 
    
    Figure \ref{fig:eads_janus}(c) shows the site and orbital projected DOS for the atoms at the interface. The orbital and site DOS for the combined system demonstrate hybridization, as was also seen by the overlap in the DOS for the 2D and bulk material. When overlaid with the DOS of the isolated 2D material (i.e. pre-adsorption), the profiles reveal that once adsorbed, \sample\ undergoes a semiconductor-to-metal transition and/or hybridizes with the metallic states of the bulk material (see SI Figure 5). Even for cases such as the \sample\ on the Sr-(111) bulk surface where the $z$-separation distance is large ($\sim$3.1 \AA), it is still enough interaction for the adsorbed DOS profile to change appreciably from that of the isolated 2D material. This hybridization effect, as well as the semiconductor-to-metal transition, reveals the significant impact that the choice of bulk support can have on 2D material properties through 2D-bulk interactions.

\begin{figure}[t]
        \centering
        \includegraphics[width=6.66in]{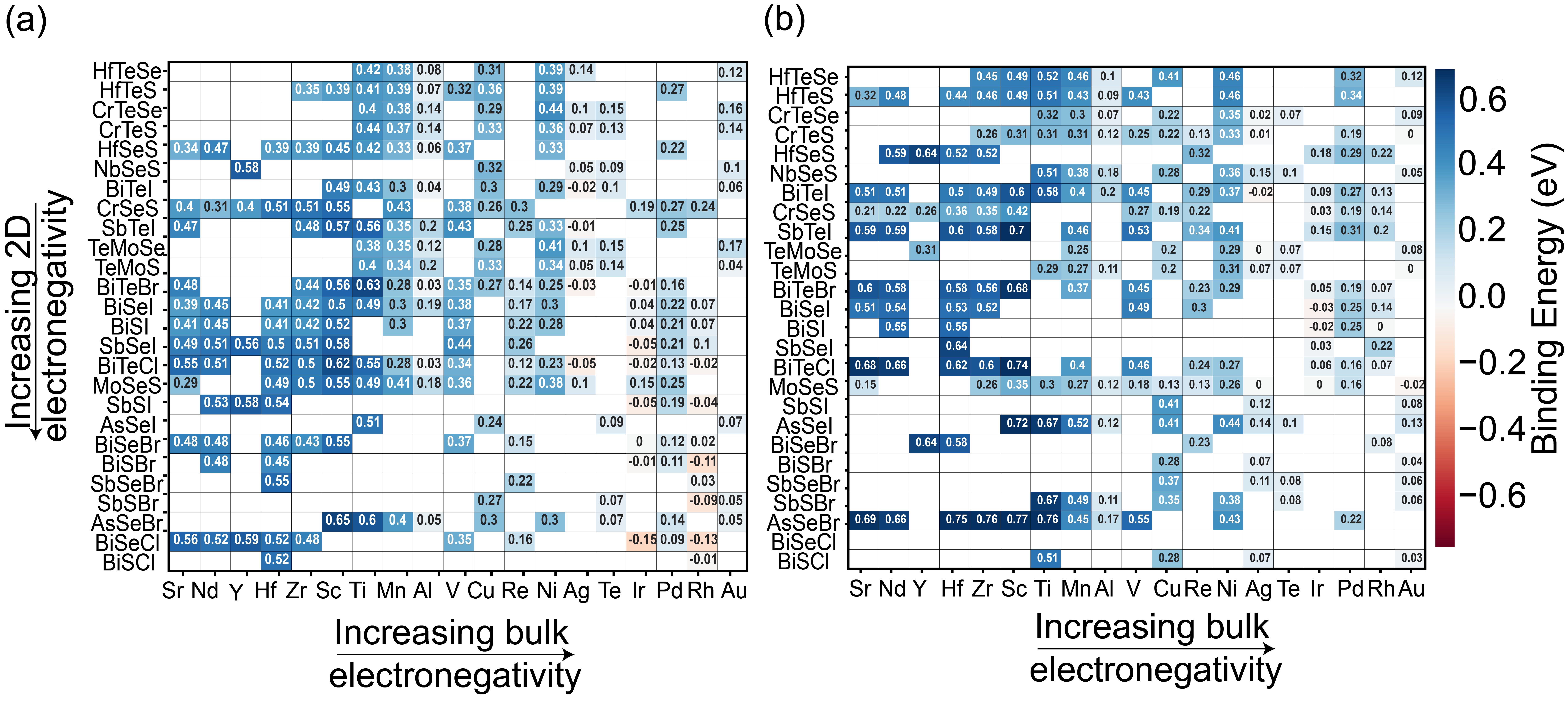}
        \caption{Heatmap of the \bind\ for the (a) $1T$ and (b) $1H$ phase of the Janus 2D materials with the 2D materials listed on the $y$-axis and the bulk materials on the $x$-axis and the color map represents the (\bind) binding energy. The materials are sorted based on their electronegativities.}
        \label{fig:subfigure_heatmaps}
    \end{figure}

    We further examine these potential bulk effects on all the heterostructures using heatmaps, such as those shown in Figure \ref{fig:subfigure_heatmaps}. When considering the equation for \ads\ (the difference between \form\ and \bind\ ), it is evident that \form\ is constant for any given 2D material. Thus, variations in \bind\ for a given 2D phase over a range of bulk substrates directly reflects the changes in the interaction's strength between the 2D and the bulk material. It can therefore be more useful to analyze \bind\ rather than the adsorption energy to discern trends within the data. 
     
     The heatmaps in Figure \ref{fig:subfigure_heatmaps}(a) and \ref{fig:subfigure_heatmaps}(b) show how \bind\ varies for a given 2D material over all of its various bulk substrate surfaces (left to right), and vice versa (top to bottom). The materials on the axes of the \bind\ heatmaps are sorted from least (top-left) to greatest (bottom-right) electronegativity of the 2D and bulk materials (electronegativies were obtained from the $Matminer$'s database.) 
     
    In these heatmaps, it can be seen  that \bind\ changes much more drastically moving from left to right (i.e. changing the bulk materials for a fixed 2D material) than it does moving top to bottom (i.e. changing the 2D material for a fixed bulk support).  In addition, the \bind\ appears to have a strong correlation with the difference in electronegativities of the 2D and bulk material (e.g. highest \bind\ in the bottom-left corner, with the lowest bulk and highest 2D electronegativities). In the following section on ML models, we further examine such trends through the percent importance of various features in predicting the \bind\, including electronegativity differences, along with other numerous distinct properties of the 2D and bulk materials.

\subsubsection{Distribution of 2D-Substrate $Z$-Separation Distances}
   
\begin{figure}[h!]
        \centering
        \includegraphics[width=6.66in]{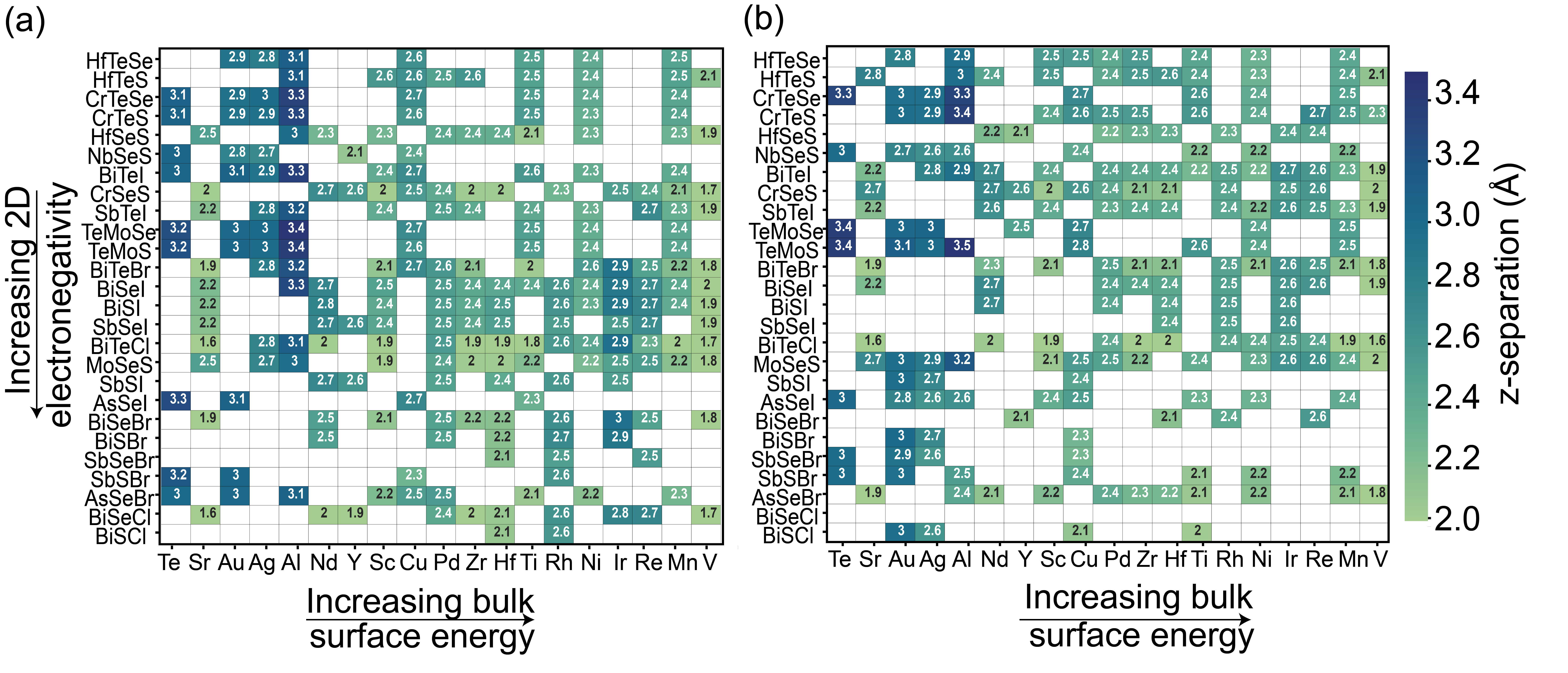}
        \caption{Heatmap of the $z$-separation for the (a) $1T$ and (b) $1H$ phases of the Janus 2D materials, sorted based on the electronegativity of the 2D materials ($y$-axis) and  bulk's surface energy ($x$-axis). The color map represents the $z$-separation.}
        \label{fig:subfigure_heatmaps2}
    \end{figure}

    Similarly, Figure \ref{fig:subfigure_heatmaps2} (a) and (b) contain the heatmaps for $z$-separation for both $1T$ and $1H$ phases of the 2D-bulk heterostructures. For the $z$-separation heatmaps, the 2D materials on the $y$-axis are still sorted based on their electronegativities, while the bulk materials on the $x$-axis are sorted based on their surface energies. It can also be seen that just as in \bind\, a change in the  bulk support's surface energy (moving left to right) yields a greater spread of $z$-separations than changing the 2D material (moving top to bottom). In both polytypes the highest $z$-separation corresponds with the lowest 2D electronegativities and lowest bulk surface energies (top-left corner). 

    The clear correlation between \bind\ and $z$-separation, and consequently \ads\ and the $z$-separation, suggests that, in principle, it may be possible to tune the properties of the 2D material by simply changing the bulk support. 
    
    As the support is changed, the $z$-separation could then be effectively modulated by in turn altering the bonding and interaction strength between the 2D and bulk material. However, to utilize this trend in device applications, a clear set of guidelines needs to be established regarding the underlying material-dependent properties that give rise to the variations in interaction strength for any given 2D-bulk material pair and the consequent impact on the 2D material.

\subsection{Machine Learning Insights into the Fundamental Factors Governing Janus 2D Heterostructure Stability}
    As discussed above, the thermodynamic stability of the heterostructures, i.e. \ads, and therefore the \bind, varies when the bulk support is changed. However, the factors that govern the thermodynamic stability of the heterostructures of 2D and bulk materials are not directly discernible from the data plotted in Figures \ref{fig:subfigure_heatmaps} and \ref{fig:subfigure_heatmaps2}. 
    
    In this section, we use ML to further explore the factors that influence the thermodynamic stability of the heterostructures. We develop two ML models for predicting the target heterostructure properties- \bind\ and $z$-separation - from the dataset of the highest \bind\ heterostructure configurations, i.e the most stable configuration for each 2D-bulk pair. A total of 438 2D-bulk pairs are obtained. We use the features that are discussed in the Machine Learning Methodology section. The predictive capabilities of the ML models and the most important features that dictate the \bind\ and $z$-separation are discussed below.

\subsubsection{ML Model for Determining the Binding Energy of 2D-Bulk Material Heterostructures}

   \begin{figure}[h!]
            \centering %
            \includegraphics[width=6.66in]{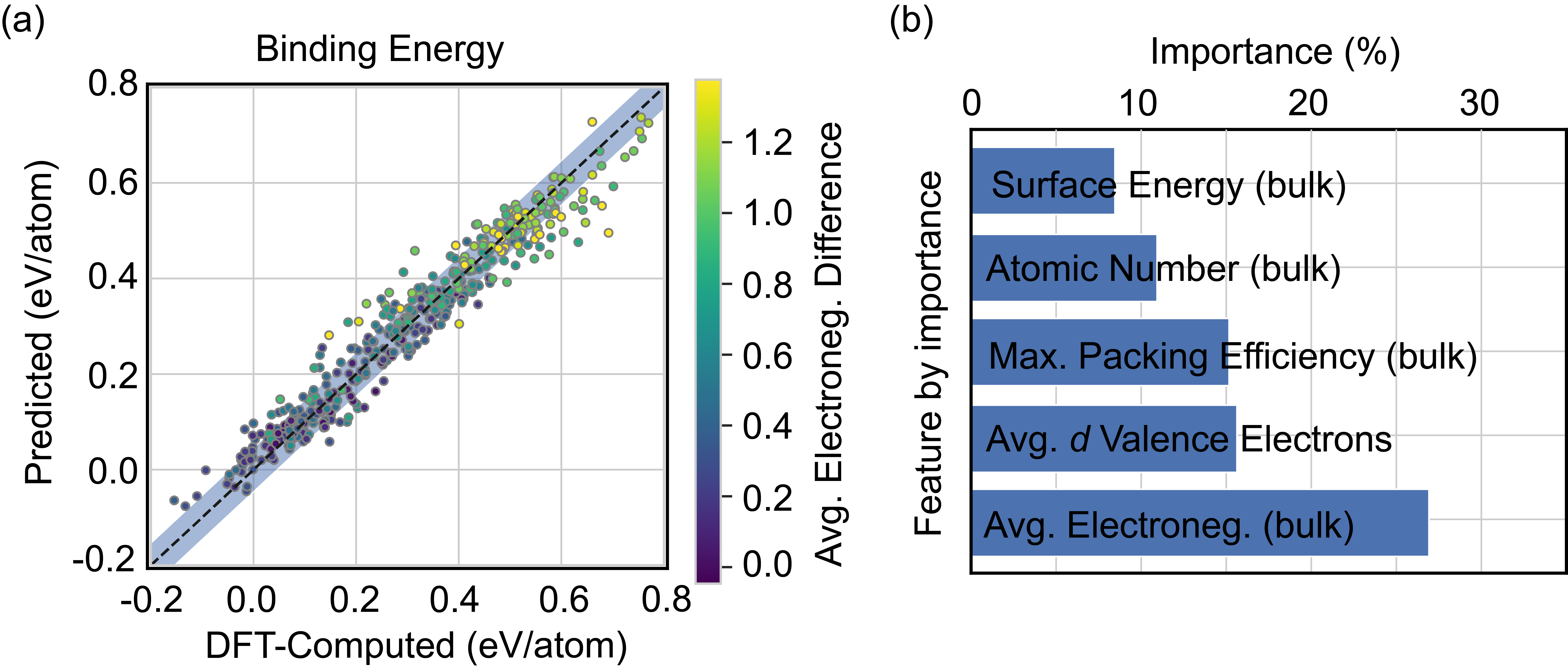}
            \caption{(a) Scatter plot compares the DFT-computed and machine learning-predicted \bind~for the Janus 2D-bulk heterostructure configurations. The color map represents the value of the electronegativity difference between the 2D material and bulk material. The black dashed line indicates the $y=x$ function. The closer the data points are to this line, the better the agreement between the predicted \bind\ value to the DFT computed value. The shaded region is the RMSE provided by the model. (b) The top five most predictive features of the ML model are shown. 4 of the 5 top features are dictated by the bulk substrate properties.}
            \label{fig:mlbind}
        \end{figure}

The RFR$^{bind}$ model, trained on the target property \bind\ , yielded good accuracy, with R$^2$= 0.940, a mean absolute error (MAE) of 0.036 eV/atom, and root mean squared error (RMSE) of 0.047 eV/atom. Figure \ref{fig:mlbind}(a) shows the DFT-computed and ML-predicted \bind\ for RFR$^{bind}$, where the shaded gray area represents the RMSE from the cross-validation, and the $y=x$ dashed line indicates optimal matching between the DFT and ML-predicted values. The ML-predicted binding energies closely follow this $y=x$ dashed line, indicating the model's accuracy in predicting the \bind\ of the 2D-bulk heterostructure pairs, which is further confirmed by the low MAE and RMSE values.

\begin{table}
\caption{Summary of machine learning statistics of random forest regression (RFR) models trained on either binding energy or $z$-separation as target properties, where R$^2$ is the coefficient of determination, RMSE is root mean squared error, and MAE is the mean absolute error. RFR$^{bind}$ is the model excluding electronegativity difference feature, while RFR$^{bind'}$ includes it. RFR$^{z-sep}$ is the model excluding $\delta_{2D}^z$, the change in 2D thickness, while RFR$^{z-sep*}$ includes it. }
\centering
    \begin{tabular}{|c|c c|c c|}
    \hline
      & RFR$^{bind}$ & RFR$^{bind'}$ & RFR$^{z-sep}$ & RFR$^{z-sep*}$\\
                                    \hline
    R$^2$ & 0.940  & 0.937 & 0.838 & 0.886\\
    RMSE & 0.047 eV/atom & 0.048 eV/atom & 0.139 \AA & 0.118 \AA \\
    MAE & 0.036 eV/atom &  0.037 eV/atom & 0.095 \AA & 0.084 \AA \\ 
    \hline
    \end{tabular}
    \label{tab:ml_stats}
\end{table}

From the physics of bonding science, one would intuitively expect that the higher the electronegativity difference between the 2D and bulk material, the more stable the heterostructure, and the higher its \bind\ . Electronegativity difference is often a strong determinant of bonding stability. The \bind\ data in Figure \ref{fig:mlbind}(a) are colored with the 2D-bulk electronegativity difference feature from the data set, and a strong correlation between this feature and the \bind\ is clearly visible, with the expected relationship. The relative importance of each feature in predicting \bind\ can be extracted from the RFR model after training. While 2D-bulk electronegativity difference was indeed output as one of the highest-importance features (second-highest, 13 \% importance), the highest-importance feature extracted was actually the bulk electronegativity, with a significant jump in relative importance (29 \%importance). 

To explore this further, we excluded the electronegativity difference as a feature from the ML training and found that the resultant RFR model has a slightly better performance--MAE of 0.036 eV/atom instead of 0.037 eV/atom (statistics summarized in Table \ref{tab:ml_stats}, where RFR$^{bind'}$ is the initial model with difference included). The updated RFR$^{bind}$ model excluding the electronegativity difference feature is the one displayed in Figure \ref{fig:mlbind}, with electronegativity of the bulk retaining its highest-importance placement among the features (see Figure \ref{fig:mlbind}(b)). This dominating effect of bulk electronegativity as compared to electronegativity difference can potentially be explained in relative electronegativity ranges of our sampled materials. The Janus 2D materials in our study all had electronegativities falling between 1.98-2.59, which is a much narrower range than that of the study's bulk materials (0.95-2.54). Thus, for this subset of materials, the bulk electronegativity plays a much larger role than the difference in determining heterostructure stability (i.e. \bind\ ). While it is likely that bulk electronegativity may similarly dominate important features for heterostructures of other 2D materials classes, it would be beneficial to study 2D materials with a much larger range in electronegativities to likewise establish their role in the thermodynamic stability of heterostructures.

Of the top five most predictive features for the RFR$^{bind}$ model, as shown in Figure \ref{fig:mlbind}(b), four are dependent on the bulk substrate. Aside from bulk electronegativity, the remaining high-importance features of RFR$^{bind}$ are surface energy of the bulk, structural factors of the bulk (atomic number, packing efficiency), and average 'd' valence electrons of the heterostructure. This higher dependence on the bulk is physically intuitive, as the bulk surfaces are formed by breaking stronger metallic bonds within the elemental bulk material, whereas the 2D Janus materials are created by breaking the much weaker vdW bonds of their bulk counterparts. Thus, the bulk materials, with high-energy broken bonds at their surface, will contribute much more to the heterostructure's energetics than the 2D materials, which have only weak vdW’s bonds broken at their surface.

The determination of heterostructure stability by the bulk substrate is something that has already been observed experimentally for growth of transition metal dichalcogenides (TMDCs)~\cite{geng2018recent, siegel2019growth, loh2014growth, lee2012synthesis, novoselov2012roadmap}. Some bulk substrates are known for enabling successful growth for a wide number of TMDCs (e.g. Cu is used for growth of MoS$_2$, WS$_2$, MoSe$_2$, NbS$_2$, etc.) whereas other substrates universally lead to unsuccessful growth of TMDCs (e.g. the Ni substrate). The ML feature analysis in this work reveals that the thermodynamic stability of 2D material-bulk heterostructures can thereby be effectively controlled by tweaking the properties of only the bulk material.  

\begin{figure}[!t]
            \centering
            \includegraphics[width=6.66in]{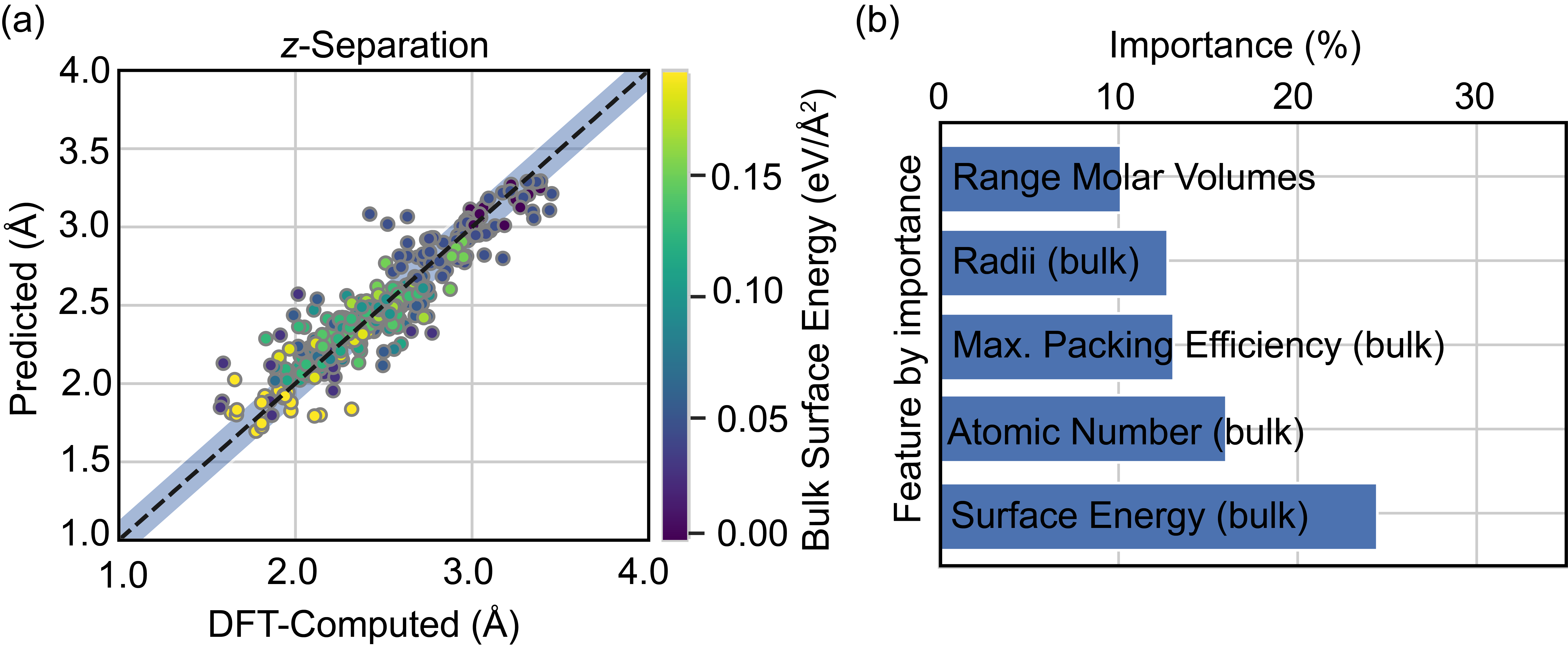}
            \caption{(a) Scatter plot comparing the DFT-computed and machine learning-predicted $z$-separation distance. The color map represents the value of the bulk's surface energy. The black dashed line indicates the $y=x$ function. The shaded region is the RMSE provided by the model. (b) The bar chart indicates the top five most predictive features return by the random forest regression model for predicting the $z$-separation distance between the Janus 2D materials and the bulk surface.}
            \label{fig:mlzsep}
        \end{figure}

To further gain insights on the interfacial structure of 2D-bulk heterostructures, a separate predictive RFR model was trained, this time with the $z$-separation as the target property. The final two columns of Table \ref{tab:ml_stats} compare two versions we studied of the RFR $z$-separation model -- RFR$^{z-sep}$ and RFR$^{z-sep*}$. The initial 11-feature model RFR$^{z-sep*}$ model yields a reasonably good accuracy, with an R$^2$ of 0.886 and MAE of 0.084 \AA. When considering the 5 highest-importance features extracted from this model (list shown in SI Figure 4(b)), four of the five highest-importance features are readily available without prior computational and/or experimental determination, either from existing materials databases (bulk surface energy) or from intrinsic material properties (bulk atomic number, maximum packing efficiency, and bulk radii). The remaining feature, $\delta_{2D}^z$, is the change in the adsorbed 2D material's thickness in the $z$-direction, and is obtained in this study after relaxing the 2D-bulk heterostructure through DFT. 

In order to determine if a similar predictive ML model can be yielded with only non-DFT-computed material properties, we retrained the $z$-separation model with the $\delta_{2D}^z$ feature excluded (shown in Figure \ref{fig:mlzsep}. While this RFR$^{z-sep}$ model yielded lower accuracy (R$^2$ = 0.838, MAE= 0.095 \AA), it has the advantage of yielding reasonable $z$-separation predictions while relying mainly on features which are readily known or easy to obtain. As seen in Figure \ref{fig:mlzsep}(b), the high-importance feature now replacing $\delta_{2D}^z$ is the range of elemental molar volumes, another input with no additional computational cost.  

Just as for the \bind\ model, the highest-importance feature for RFR$^{z-sep}$ (bulk surface energy) is plotted on the color map in Figure \ref{fig:mlzsep}(a). Interestingly, as the surface energy increases, the $z$-separation distance between the 2D material and bulk surface decreases. This can be attributed to the larger number of dangling bonds present at the surface of these higher-surface-energy bulk materials, increasing the strength of the 2D and bulk material's interaction. This RFR$^{z-sep}$ model, in combination with the predictive binding energy model RFR$^{bind}$, can be used to identify 2D-bulk pairs that have both suitable binding energies and optimal $z$-separation distances for tuning the 2D-bulk interactions. More specifically, the RFR$^{z-sep}$ model in this work can be used with exclusively non-calculated features, potentially allowing for an easily extendable model for other unstudied 2D-bulk systems, while also demonstrating the feasibility of developing more low-cost ML models in future for predicting heterostructure stability. Such models can thus be used to identify promising bulk supports for both stabilizing and tuning a given 2D material, thus potentially allowing for the tailoring of stable and effective 2D-bulk heterostructures for a range of applications such as  nanoelectronics, sensing, and energy conversion.

\section{Conclusion}
    In summary, the complex interfaces of nearly 1000 heterostructures of 2D Janus and bulk materials' surfaces were investigated using \emph{ab initio} simulations and machine learning. Through these calculations, clear interfacial trends, such as correlation between higher binding energies and lower $z$-separations, were identified. Random forest regression ML models were then trained on the most stable configurations ($\sim$400 heterostructures), in order to predict target properties such as binding energy and $z$-separation. These ML models provided critical insights on 2D-bulk interactions, demonstrating that the bulk support largely governs stability and bonding in the 2D-bulk heterostructure. In both binding energy and $z$-separation models, the most important features in prediction were also both found to be bulk properties (i.e. bulk electronegativity and surface energy, respectively). The results of these calculations were compiled in an accessible new database, aiHD, (https://hydrogen.cmd.lab.asu.edu/data)for $>$1200 2D-bulk heterostructures and their associated properties.

\section{Data Availability}
    The data underlying this study are freely available on the \href{http://hydrogen.cmd.lab.asu.edu/data}{aiHD} website and by the authors upon reasonable request.
\begin{acknowledgement}

This work was primarily supported by the U.S. Department of Energy, Office of Science, Basic Energy Sciences under Award \# DE-SC0024184 (machine learning). This work was also supported by start-up funds from Arizona State University ($ab$ $initio$ simulations). R. G. acknowledges a graduate fellowship through the National Science Foundation Graduate Research Fellowship Program under Grant No. 026257-001 from August 2020 to August 2023. This work used the Extreme Science and Engineering Discovery Environment (XSEDE), supported by National Science Foundation grant number TG-DMR150006. The authors acknowledge Research Computing at Arizona State University for providing HPC resources that have contributed to the research results reported within this paper. This research also used resources of the National Energy Research Scientific Computing Center, a DOE Office of Science User Facility supported by the Office of Science of the U.S. Department of Energy under Contract No. DE-AC02-05CH11231. The authors acknowledge Akash Patel for his dedicated work maintaining our database and API.

\end{acknowledgement}

\begin{suppinfo}
The supporting information provides a detailed description of the 2D materials, the bulk materials, feature selection, and the machine learning model training and validation details.

\end{suppinfo}

\bibliography{references}

\end{document}